\documentclass[useAMS,usenatbib]{mn2e}
\usepackage{natbib}
\usepackage{color,graphicx} % NB you need to use graphicx or else you can't use includegraphics command
\usepackage{amsmath}
\usepackage{amsfonts}
\usepackage{amssymb}
\usepackage{wasysym}

\author[Keane \& Kramer]{E.F.~Keane \& M. Kramer \\
University of Manchester, Jodrell Bank Centre for
Astrophysics Alan-Turing Building, Oxford Road, Manchester M13 9PL, UK}
\date{8 October 2008} 
\title[On the birthrates of Galactic neutron stars]
{On the birthrates of Galactic neutron stars}

\begin{document}

\maketitle

\begin{abstract}
  In light of the recently discovered neutron star populations we discuss the
  various estimates for the birthrates of these populations. We revisit the
  question as to whether the Galactic supernova rate can account for all of
  the known groups of isolated neutron stars. After reviewing the rates and population 
  estimates we find that, if the estimates are in fact accurate, the current birthrate 
  and population estimates are not consistent with the Galactic supernova rate.  
  We discuss possible solutions to this problem including whether or not some 
  of the birthrates are hugely over-estimated. We also consider a possible evolutionary
  scenario between some of the known neutron star classes which could solve
  this potential birthrate problem. 
\end{abstract}


\section{Introduction}         % introduction

In the standard scenario, neutron stars (NSs) are formed during the core
collapse of massive stars which links their number in the Galaxy to the
Galactic supernova rate. The number of Galactic NSs can be inferred from
observations, taking the various manifestation of NSs into account. In recent
years, new and different observational manifestions of NSs have been
discovered, so that it is warranted to study the impact, if any, of these
discoveries onto the birthrates that is required to sustain this increased NS
population.

The new manifestations of NSs include Rotating Radio Transients (RRATs;
McLaughlin et al.~2006)\nocite{mll+06} and X-ray Dim Isolated Neutron Stars
(XDINS; see Haberl 2007 and references therein)\nocite{hab07}. These objects
join the $\sim1800$ known radio pulsars and the small group of
magnetars~\citep{wt04}.  Do these previously unknown types of observable NSs
increase the overall population by an amount that it is difficult to reconcile
the formation rates with those predicted by theory? The basic requirement we
make to answer this question is that the individual birthrates of the
different NS populations should not exceed the Galactic core-collapse
supernova (CCSN) rate, i.e.
\begin{equation}\label{balance_equation}
	\beta_{{\rm CCSN}}\ge\beta_{{\rm total}}=\beta_{{\rm PSR}}+\beta_{{\rm XDINS}}+\beta_{{\rm RRAT}}+\beta_{{\rm magnetar}},
\end{equation}
where $\beta_{{\rm X}}$ is the birthrate (per century) of a NS of type X.

Recently, this question has also been addressed by~\citet{ptp06} where it
was concluded that this requirement can be met if we assume that XDINSs are
in fact nearby RRATs. However, as we detail below, the pulsar birthrate
considered is a lower limit which has since been superceded. In addition,
the recent non-detection of any radio RRAT-like bursts from the
XDINSs~\citep{kba+08} means that the identification of these two populations
is not certain.  Furthermore recent work suggests that the heretofore
neglected magnetar contribution may not be negligible so that the question
as to whether the CCSN rate requirement is satisfied is reinstated.

The aim of this paper is to study the posed question by investigating the most
recent knowledge about each contributing NS population and its Galactic
birthrate. After introducing each manifestation of NS in turn, we will revisit
the estimates for all terms in Eqn.~\eqref{balance_equation}. The results are
then discussed in detail before conclusions are drawn.

\section{Different manifestations of neutron stars}

\subsection{Radio pulsars}
\label{sec:pulsar}

Radio pulsars are rapidly rotating, highly magnetised NSs.  Coherent radio
emission is produced by a pair plasma above the magnetic polar caps of the NS,
believed to orginiate from particle cascades after an acceleration of
electrons and positrons in the strong electric and magnetic fields
(e.g.~Lorimer \& Kramer 2005). The spectra for this emission typically
increases with decreasing radio frequency with mean spectral index of
$-1.8$~\citep{mkk+00} before peaking in the range $100-300$
MHz~\citep{mgj+94}.  Pulsar periods range from 1.4 ms up to 8.5 s with two
distinct distributions - the ``normal'' radio pulsars which have periods of
$\sim500$ ms and the so-called `millisecond pulsars' with typical periods of
$\sim5$ ms. Figure 1 shows a $P-\dot{P}$ diagram, a standard pulsar
classification tool, where these two populations are easily identified. The
standard model of pulsar physics assumes pulsars have dipolar magnetic fields
and that the loss of rotational energy powers the pulsar. With these we can
determine the ``characteristic surface magnetic field'' of the pulsar to be
\begin{equation}
	B_{{\rm S}}=3.2\times10^{19} {\rm G} \sqrt{P\dot{P}},
\end{equation}
which assumes canonical NS values of $M=1.4~{\rm M_{\astrosun}}$, $R=10$ km, 
a moment of inertia of $I=(2/5)MR^2=10^{45}\mbox{\rm g cm}^2$ and an orthogonal rotator 
(e.g.~Lorimer \& Kramer 2005).\nocite{lk05} Assuming a spin-down law of the form
$\dot{P}=KP^{2-n}$ and that pulsars are born spinning much faster than what
  we currently observe (i.e. $P_{{\rm birth}}\ll P_{{\rm now}}$) we can
  determine a spin-down age of
\begin{equation}
	\tau=\frac{1}{(n-1)}\frac{P}{\dot{P}}.
\end{equation}
For a pure dipole, the `braking index' $n=3$, resulting in the ``characteristic
age'' $\tau_{{\rm c}}=P/2\dot{P}$. We can also define the pulsar ``spin-down
luminosity'' $\dot{E}=-d/dt[(1/2)I\Omega^2]$. Assuming again the canonical NS
values, we find
\begin{equation}
	\dot{E}=3.95\times10^{31} \; \mbox{\rm ergs s}^{-1} \;
	\left( \frac{\dot{P}}{10^{-15}} \right) \;
	\left( \frac{P}{\mbox{\rm s}} \right)^{-3}.
\end{equation}
Lines of constant $\dot{E}$, $B_{{\rm S}}$ and $\tau_{{\rm c}}$ are shown in
Figure 1 along with different evolutionary paths for different braking
indices. The lower right area of the diagram devoid of any pulsars is known as
the pulsar `death valley'~\citep{cr93a}. It is here that it is believed that
the electric potential at the polar caps is insufficient for ripping particles
from the NS surface, hence failing to provide the plasma needed for radio
emission.

\begin{figure}
	\begin{center}
		\includegraphics[scale=0.25]{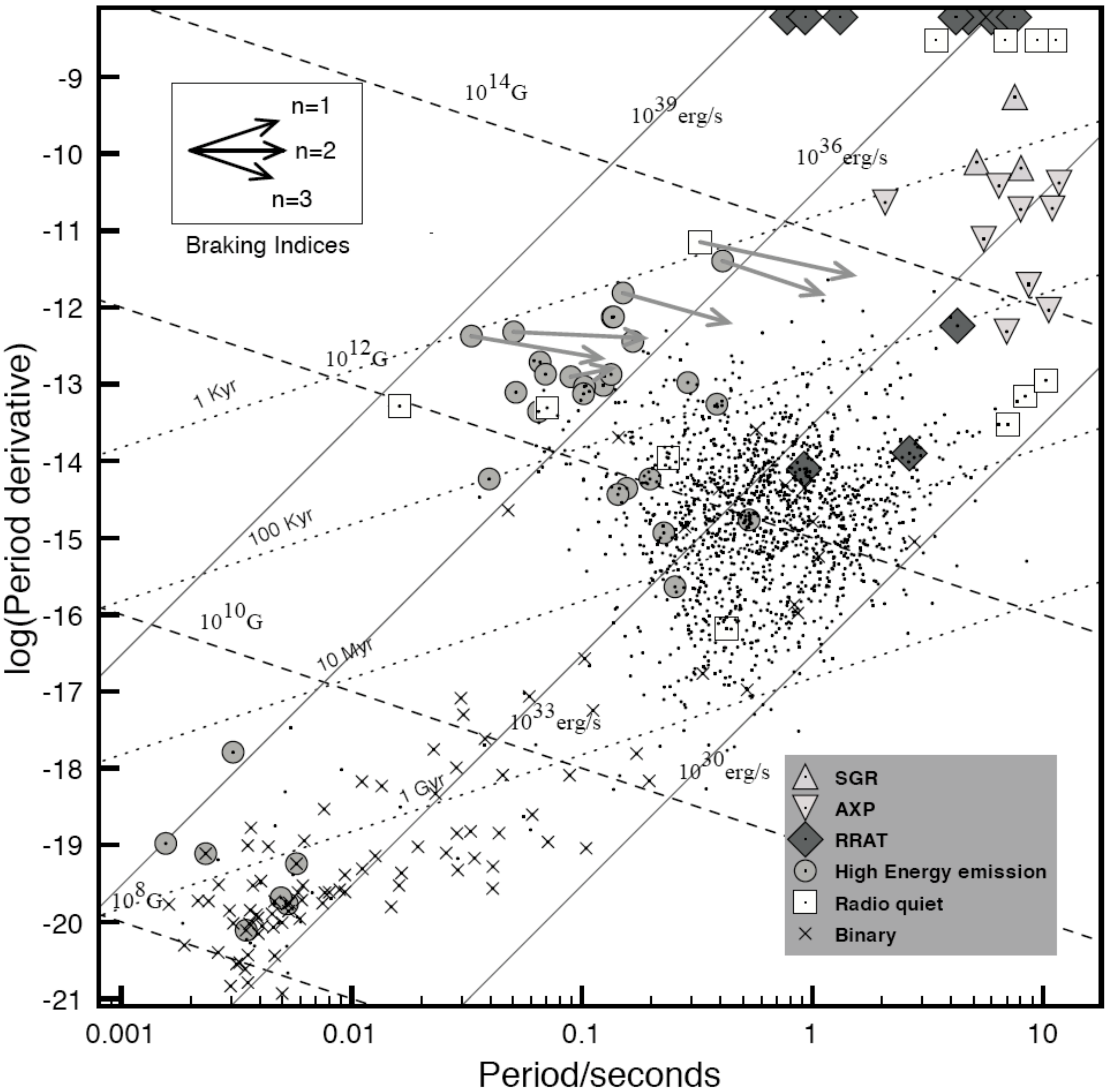}
	\end{center} 
	\caption{A $P-\dot{P}$ diagram showing the various NS populations. The 4 XDINSs and 8
		RRATs without a known $\dot{P}$ are placed at the top
		right at their respective periods. Arrows indicate the
		current spin evolution of those sources for which a
		braking index has been measured (Figure provided by
		C. Espinoza).} 
	\label{fig:ppdot}
\end{figure}

\subsection{Millisecond Pulsars and X-ray Binaries}

The standard evolutionary picture for millisecond pulsars (e.g.~Alpar et
al.~1982)\nocite{acrs82} is that they are born in supernovae with periods of
10s of milliseconds, then evolve along a line of approximately constant
magnetic field strength (i.e. $n=3$) on the $P-\dot{P}$ diagram, slowing down
until eventually radio emission ceases once they pass into the pulsar death
valley. Here, those `dead' pulsars which happen to be in binary systems can
undergo accretion from their binary companion. This accretion can heat areas
of the NS surface (`hot spots') which emit strongly in X-rays - the system is
now an X-ray binary system. As well as heating the star the accretion can spin
up the star to periods of a few milliseconds.  The pulsar is now reborn as a
millisecond pulsar and once again is seen to emit as a radio pulsar. In what
follows we do not consider the NSs which are millisecond pulsars or those seen
in X-ray binaries as this standard evolutionary picture sees these two
populations as originating from `normal' radio pulsars. Their birthrates
should thus be accounted for in the pulsar birthrate.  However we note that if
some NSs in X-ray binaries did not originate from the normal radio pulsars the
problem outlined below may be emphasised even further.

\subsection{RRATs}           % RRATs - what are they - basic properties ...
\label{sec:rrat}

The discovery of eleven RRAT sources was made (McLaughlin et al. 2006)~\nocite{mll+06} from single pulse
searches of the 1.4-GHz Parkes Multi-beam Pulsar Survey
(PMPS) (Manchester et al. 2001)~\nocite{mlc+01}. These are distant ($\sim2-7$ kpc), transient sources
which emit single radio bursts. The burst arrival times are stochastic in some
sources with others showing distinct on-off states (M. A. McLaughlin, private
communication). The times between bursts range from 4 min to 3 hr with typical
burst rates $\dot{\chi}\sim1$ ${\rm hr^{-1}}$. The bursts are narrow ($2-30$
ms) and rather bright (with peak flux densities of $0.1-3.6$ Jy) and thus have
high brightness temperatures of $10^{22}-10^{23}$ K. These values are higher
for the RRATs than all known sources except for the giant radio pulses and
nano-giant pulses emitted by the Crab pulsar which have $T_{\rm  B}\approx 
10^{31}$ K~\citep{hr75} and $T_{{\rm B}}\approx10^{38}$
K~\citep{hkwe03}, respectively.

Continued observations of these sources have enabled underlying periodicities
to be determined in all eleven sources and period derivatives to be determined
for three sources. Periods in the range $0.7-7$ s are seen which suggests a NS
origin. This seems to be confirmed from X-ray observations of the most
prolific source - RRAT J1819$-$1458. Using first Chandra (Reynolds et al. 2006)~\nocite{reyn+06} and
then XMM Newton (McLaughlin et al. 2007)~\nocite{mrg+07} a thermal X-ray spectrum (characteristic of a
cooling NS) with $kT\sim140$ eV was found, as well as X-ray pulsations at the
identical period as determined from the radio observations.

We can place the three sources with known period derivative on the $P-\dot{P}$
diagram and estimate their surface magnetic field strength using Eqn.~(2).
The magnetic field strengths are in the range of the normal radio pulsars
($\sim{\rm few\times10^{12}}$ G) except for J1819$-$1458 which lies between
the normal pulsars and the magnetars with $B_{{\rm S}}=5\times10^{13}$ G. The
discovery of a spectral feature in the X-ray spectrum of J1819$-$1458 (which
may be due to proton cyclotron resonant scattering) seems to support this
estimate~\citep{mrg+07}.

\subsection{XDINSs}            %XDINSs
\label{sec:xdins}

The XDINSs are a small group of radio-quiet, close-by ($\sim100$ pc) X-ray
pulsars situated in the Gould Belt, a local, partial ring of stars which includes the sun (Poppel, 1997\nocite{poppel97}). 
XDINSs were originally discovered over a
decade ago~\citep{wwn96} with seven sources now known (sometimes referred to
as ``The Magnificent Seven''). XDINSs have thermal X-ray spectra with
${\rm kT=50-100}$ eV and show X-ray pulsations with periods in the range $\sim3-11$
s~\citep{hab04}. All seven sources have determined periods with three well known period 
derivatives and upper limits for three more (see Tables 1 \& 3 in 
Haberl 2007, Tiengo \& Mereghetti 2007, van Kerkwijk \& Kaplan 2008)\nocite{tm07,kk08}. However the upper limits 
determined are $1-2$ orders of magnitude higher than the three well known $\dot{P}$ 
values so may not be very constraining.
We can place the three sources with known $\dot{P}$ on the $P-\dot{P}$ diagram
and can determine $B_{{\rm S}}\sim10^{13}$ G in the standard way. These 3 sources lie just below the magnetars. 

The X-ray spectra of XDINSs can be fit well with a single blackbody and interestingly do not require a power-law 
component which suggests that XDINSs do not have magnetospheres. 
Also, as for RRATs there are observed spectral features which may be due to proton-cyclotron lines
in a strong magnetic field~\citep{hab07}. We note that there is much
current work underway, searching for RRAT-like bursty emission from XDINSs.
However, no emission has been found above a flux density limit of
$\sim10$ ${\rm \mu Jy}$ (Kondratiev et al. 2008)\nocite{kba+08} 
from 820-MHz observations with the Green-Bank
Telescope. In addition there has been no detection with GMRT at 320 MHz (B. C. Joshi, private communication). 
Searches are also underway using the Parkes 
telescope at 1.4 GHz (A. Possenti, private communication).
These non-detections suggest that the identification of XDINSs as nearby RRATs~\citep{ptp06} 
might be incorrect.

\subsection{Magnetars}      %Magnetars
\label{sec:magnetar}

It is thought that both Soft Gamma Repeaters (SGRs) and Anomalous X-ray
Pulsars (AXPs) belong to the magnetar class of NSs~\citep{wt04}. Magnetars are
believed to be isolated X-ray pulsars with strong magnetic fields
($10^{14}-10^{15}$ G) and periods in the range $2-12$ s and were, until
recently, thought to be radio-silent. However transient radio emission
has been detected from two AXPs - XTE J1810$-$197~\citep{cmh+06} and 1E
1547.0$-$5408 (Camilo et al. 2008)\nocite{crj+08} with both sources showing flat radio spectra. This is different to what is
seen in normal radio pulsars (see Section~\ref{sec:pulsar}).
Magnetic fields inferred again from the observed spin and spin-down rates are
shown for magnetars in the same $P-\dot{P}$ diagram in Figure~\ref{fig:ppdot}.

\section{Birthrates}

\subsection{The Core-Collapse Supernova Rate}     % What Diel et al got

Recently, the CCSN rate (for Type Ib, Ic and Type II SNe) has been
determined from measurements of $\gamma$-ray radiation from $^{26}$Al in the
Galaxy (Diehl et al. 2006)~\nocite{diehl+06}. Quantifying this $\gamma$-ray emission allowed the
authors to weigh the amount of $^{26}$Al in the Galaxy, as each CCSN expels a
well known yield of $^{26}$Al.
Assuming an initial mass function\footnote{The IMF,
  $\Phi(m)dm$, is the probability that a star will be born with an initial
  mass in the range of $m+dm$.}  (IMF), a Scalo IMF
(${\rm {dlog}\Phi/{dlogm}=-2.7}$ for high masses),
they inferred the Galactic CCSN rate to be
\begin{equation}
	{\rm \beta_{CCSN}=1.9\pm1.1 /century}.
\end{equation}

% Can work it out to check 
As a consistency check, we integrate the IMF to compute,
\begin{equation}
	\beta_{\rm CCSN}=\frac{SFR}{\langle m \rangle}f_{\rm CCSN},
\end{equation}
where $\langle m \rangle$ is the mass expectation value and $f_{CCSN}$ is the
fraction of stars which end their lives in a CCSN, i.e.~those with initial masses
in the range $(11\pm1)-25$ ${\rm M_{\astrosun}}$(Heger et al. 2003;
Podsiadlowski et al. 2004)\nocite{hfw03}. Using the ``standard'' IMF, considered to be that
of Kroupa (2001, 2002; McKee \& Ostriker 2007)~\nocite{kroupa01}~\nocite{mo07} 
and defined between $0.1-120$ ${\rm M_{\astrosun}}$, and assuming a star 
formation rate of ${\rm SFR=4M_{\astrosun}.yr^{-1}}$ (Diehl et al. 2006; Stahler \& 
Palla 2004)~\nocite{diehl+06,sp04}, we determine a CCSN rate of as high as 
$\approx1.9\pm0.9$/century if we use a Salpeter-like high mass IMF 
(${\rm {dlog}\Phi/{dlogm}=-2.3}$). While this is consistent with the rate of Diehl et al., it 
has been suggested that the observed Salpeter-like IMF may be artificially large at high 
masses due to the effect of unresolved binaries~\citep{kro02b}, so that a steeper 
Scalo-like IMF is more realistic in the high mass range of interest here. 
In this case we obtain a lower CCSN rate of just $\approx0.8\pm0.4$/century.
In both cases here we have assumed the error to be dominated by
the uncertainty of the SFR of up to $\sim50\%$~\citep{sp04}.

\subsection{Radio pulsars}

The most thoroughly studied NS population is that of radio pulsars. A recent
estimate of the birthrate and the number of radio pulsars in the Galaxy was
performed by Lorimer et al (2006)~\nocite{lfl+06} (L+06 from herein) using
1008 non-recyced pulsars detected in 1.4-GHz surveys using the Parkes
telescope (the PMPS and the Parkes High-Lattitude Survey). Using a pulsar
current analysis they determind population details for sources above a 1.4-GHz
radio luminosity threshold of $0.1 {\rm mJy.kpc^{2}}$. Pulsar current analyses
use the fact that the typical age of pulsars are much shorter than the age of
the Galaxy so that the pulsar population can be considered to be in a
steady-state. The flow of pulsars across the $P-\dot{P}$ diagram can thus be
considered a ``current'' obeying a continuity equation~\citep{pb81,vn81}. This
current, $J(P)$, equals the pulsar birth rate minus the pulsar death rate in
the period range $0-P$. Thus the maximum value of $J(P)$ equals $\beta_{{\rm
    PSR}}$ but as we have a flux limited sample of the pulsar population, the
maximum value of $J(P)$ provides only a lower limit to the pulsar birthrate.

The results obtained have model dependencies on the Galactic electron 
density model and on the pulsar beaming fraction model. The current best model 
for the electron density is the NE2001 model~\citep{cl02} and this was adopted 
as well as the Taurus \& Manchester (1998)\nocite{tm98} beaming model.
L+06 determine a birthrate of $\beta_{{\rm PSR}}=1.38\pm0.21$/century 
and $N_{{\rm PSR}}=155000\pm6000$. This result is consistent with the earlier 
work of Vranesevic et al. 2004~\nocite{vml+04} (V+04 from herein) which used 815 non-recycled PMPS pulsars to determine 
$\beta_{{\rm PSR}}=1.58\pm0.33$/century and ${N_{{\rm PSR}}=106600\pm11700}$ 
in this case determined above a higher threshold of $1 {\rm mJy.kpc^2}$. In both 
cases the now superceded TC93~\citep{tc93} electron density model was also 
used and for both analyses this produces lower birthrate estimates.

More recent work by~\citet{fk06} (FK06 from herein) yields a much higher
pulsar birthrate of ${\rm \beta_{PSR}=2.8\pm0.5}$/century. The approach of
this analysis is different - the authors model the birth properties of pulsars
(velocity distributions, magnetic fields and detectability in the PMPS and
Swinburne Multi-beam surveys) from the observational data before performing
Monte Carlo simulations to evolve the initial population to obtain the
observed pulsar sample.  The quoted birthrate is the average of 50 runs of
their simulations.  While it is twice as large as that provided by the pulsar
current analyses the results are entirely consistent as the pulsar current
analysis is to be interpreted as providing a lower limit to the pulsar
birthrate.

\subsection{RRATs}

The estimated number of RRATs is $N_{{\rm RRAT}}\gtrsim2-4\times10^5$
(McLaughlin et al.~2006) and therefore even higher than that of the radio
pulsar population (see Section 5 for discussion of the various parameters on
which this estimate depends). However, the determination of $N_{{\rm RRAT}}$
is obviously based on a very small sample of sources.  In order to account for
this uncertainty, we will use the following parameterisation in our
computations, i.e.~$N_{{\rm RRAT}}=\gamma N_{{\rm PSR}}$ where we take
$\gamma\sim1-3$.

It is important to realise the following caveat when interpreting this
estimate for the total number of RRATs. The fact that it appears to be larger
than that of pulsars does not necessarily imply that a NS is more likely to be
a RRAT than a pulsar. This would assume that the physical mechanisms for the
emission of the RRAT radio bursts is identical to that of regular
pulsars. There is no reason to assume this, especially as RRAT spectra are as
yet unknown. Emission criteria (which may represent certain ``active'' areas
on the $P-\dot{P}$ diagram) for RRAT and pulsar emission may be different and
the respective ``death-lines'' may also be different, so that the duration of
RRAT and pulsar emitting phases would not be the same either. Here lies the
advantage of considering birthrates (e.g. pulsar current analyses) rather than
absolute population numbers~\citep{ptp06}.

However, as we do not have a reliable age estimator for RRATs, we are forced 
to assume similar active lifetimes for RRATs and pulsars to work out birthrates 
from population estimates. We could conceivably use temperature as a measure 
age (see Section 5) but as there is just one RRAT with known temperature we follow 
Popov et al. (2006)\nocite{ptp06} who have argued that if RRATs are rotating 
NSs with pulsar-scale magnetic fields then the
active lifetime of pulsars $\tau_{{\rm PSR}}\approx N_{{\rm PSR}}/\beta_{{\rm PSR}}\sim 5\times10^6$ yr 
would be similar to that of RRATs, $\tau_{{\rm RRAT}}$. This holds provided the 
initial spin periods of RRATs and pulsars are within a factor of a few of each other. 
With the conclusion of approximately equal time-scales and $N_{{\rm RRAT}}=\gamma N_{{\rm PSR}}$,
this implies a RRAT birthrate of $\beta_{{\rm RRAT}}\approx\gamma\beta_{{\rm PSR}}$~\citep{ptp06}. 
Thus If we take $\gamma\sim2$ we have an indicative
RRAT birthrate of $\beta_{{\rm RRAT}}\sim2.8\pm1$/century considering the
pulsar current analyses, or as large as $\beta_{{\rm RRAT}}\sim5.6\pm1$/century considering the FK06 result.

\subsection{XDINSs}

The birthrate for XDINSs has recently been estimated by~\citet{gh07}. These
authors performed a population synthesis for XDINSs based on the seven XDINSs
detected in the ROSAT All-Sky Survey (Voges et al. 1999)\nocite{voges+99}. A
limiting volume for OB progenitor stars was determined and then compared to
the actual number of OB stars detected in this volume in the survey for the
relevant scalings. The authors then use an age estimate to find birthrates
from the simulated number of sources and determine $\beta_{{\rm
    XDINS}}\sim2.1\pm1$/century. The age estimate is arrived at simply from
averaging the characteristic age for the two XDINSs which then had well-known
$\dot{P}$'s ($\approx1.5$ \& $\approx1.9$ Myr), and earlier estimates for
their NS cooling ages ($\sim0.5$ Myr). The result is consistent with a recent
lower estimate of $\beta_{{\rm XDINS}}\sim1$/century made by~\citet{ptp06}
which used a NS cooling age of $\tau_{{\rm XDINS}}\approx1$ Myr.

\subsection{Magnetars}

Magnetar birthrates have typically been determined using two different methods
of age estimation, required to convert simulated populations to
birthrates. The first method uses spin-down age estimates for magnetars, as
used by Kouveliotou et al. (1998)~\nocite{kds+98} to determine an SGR birthrate of ${\rm
  \beta_{SGR}\approx0.1}$/century which we consider as a lower limit for the
magnetar birthrate~\citep{wt04}. An AXP birthrate was calculated similarly by~\citet{gh07} using the same population synthesis methods as for the XDINSs,
obtaining ${\rm \beta_{AXP}\sim0.2\pm0.2}$/century. Another very recent
determination of $\beta_{{\rm magnetar}}=0.15-0.3$/century has also just been
reported by Ferrario \& Wickramasinghe (2008). The second means by which
magnetar age estimates can be obtained, involves using ages of supernova
remnant associations of SGRs and AXPs. These have yielded slightly
smaller estimates (van Paradijs et al. 1995)\nocite{vtv95} as the supernova 
remnant ages tend to be longer than the spin-down ages resulting in a smaller birthrate.

Due to these small magnetar birthrate estimates relative to the other
populations of NSs, one might think that we can safely neglect the magnetar
contribution to Eqn.~\eqref{balance_equation}. However, we note the
possibility that if, for example, magnetars experience magnetic field decay
(as considered by~\cite{act04} and by~\cite{cgp00}) the true age is smaller
than the characteristic age. This would imply a higher birthrate, possibly as
high as $\sim2$/century for AXPs~\citep{gh07}. In addition, larger magnetar
birthrate estimates have been reported recently by Muno et
al. (2008)\nocite{muno+08}.  These authors studied 947 archival observations
from XMM Newton and Chandra. From the 7 magnetars detected they determine 
the most likely number of Galactic magnetars considering the small fraction of the sky
covered in these observations. They obtain, separately, birthrates for persistent AXPs, 
transient AXPs as well as a small contribution from SGRs yielding a large magnetar 
birthrate of $\beta_{{\rm magnetar}}=2.6^{+5.0}_{-1.5}$/century. This however assumes 
lifetimes of $10^4$ yr (see Figure 1) for each of these sub-populations. As the lifetime for 
transient AXPs is very uncertain it is possible that their lifetime is larger by an order 
of magnitude. In this case the persistent AXPs give the most reliable magnetar 
birthrate of $\beta_{{\rm magnetar}}=0.6^{+0.9}_{-0.3}$/century.

It is not clear if the question of beaming has been considered in the
estimates of Muno et al. but we should not necessarily expect magnetar
emission to be isotropic. In this case, only magentars beamed toward
us will have been detected.  However, judging from the observed pulse
shapes~(e.g.~Woods \& Thompsson 2004)\nocite{wt04}, we assume that the
beaming fraction is larger than for radio pulsars, so that the effect
of beaming may not be quite as significant as for pulsars. Nevertheless 
noting that the derived values may represent a lower limit, we proceed by 
neglecting this extra beaming factor and considering all the estimates
reviewed here we adopt a conservative magnetar birthrate of 
$\beta_{magnetar}\approx0.3^{+1.2}_{-0.2}$/century where our extended 
error bars allow for the potentially much higher values suggested by the Muno et al. study.

\begin{table*}
	\begin{center}
	\caption{Estimated birthrates in units of NSs/century for the different populations of NSs. The top row are the most likely values whereas the following rows give the lower limit pulsar current analyses for each of the pulsar current analyses.}
		\begin{tabular}{|c|c|c|c|c|c|c|}
			\hline\hline $\beta_{{\rm PSR}}$, $n_{{\rm e}}$ & PSRs & RRATs & XDINSs & Magnetars & Total & CCSN rate \\
			% Faucher-Giguere \& Kaspi
			\hline FK06, NE2001 & $2.8\pm0.5$ & $5.6^{+4.3}_{-3.3}$ & $2.1\pm1.0$ & $0.3^{+1.2}_{-0.2}$ & $10.8^{+7.0}_{-5.0}$ & $1.9\pm1.1$ \\
%			\hline
			\noalign{\smallskip}
			% Lorimer
			L+06, NE2001 & $1.4\pm0.2$ & $2.8^{+1.6}_{-1.6}$ & $2.1\pm1.0$ & $0.3^{+1.2}_{-0.2}$ & $6.6^{+4.0}_{-3.0}$ & $1.9\pm1.1$ \\
%			\hline
			\noalign{\smallskip}
			L+06, TC93 & $1.1\pm0.2$ & $2.2^{+1.7}_{-1.3}$ & $2.1\pm1.0$ & $0.3^{+1.2}_{-0.2}$ & $5.7^{+4.1}_{-2.7}$ & $1.9\pm1.1$ \\	
%			\hline		
			\noalign{\smallskip}
			%Vranesevic
			V+04, NE2001 & $1.6\pm0.3$ & $3.2^{+2.5}_{-1.9}$ & $2.1\pm1.0$ & $0.3^{+1.2}_{-0.2}$ & $7.2^{+5.0}_{-3.4}$ & $1.9\pm1.1$ \\
%			\hline
			\noalign{\smallskip}
			V+04, TC93 & $1.1\pm0.2$ & $2.2^{+1.7}_{-1.3}$ & $2.1\pm1.0$ & $0.3^{+1.2}_{-0.2}$ & $5.7^{+4.1}_{-2.7}$ & $1.9\pm1.1$ \\
			\hline 
		\end{tabular}
	\end{center}
	\label{tab:rates}
\end{table*}

\begin{figure}
	\vspace{-20pt}
	\begin{center}
		\includegraphics[scale=0.3]{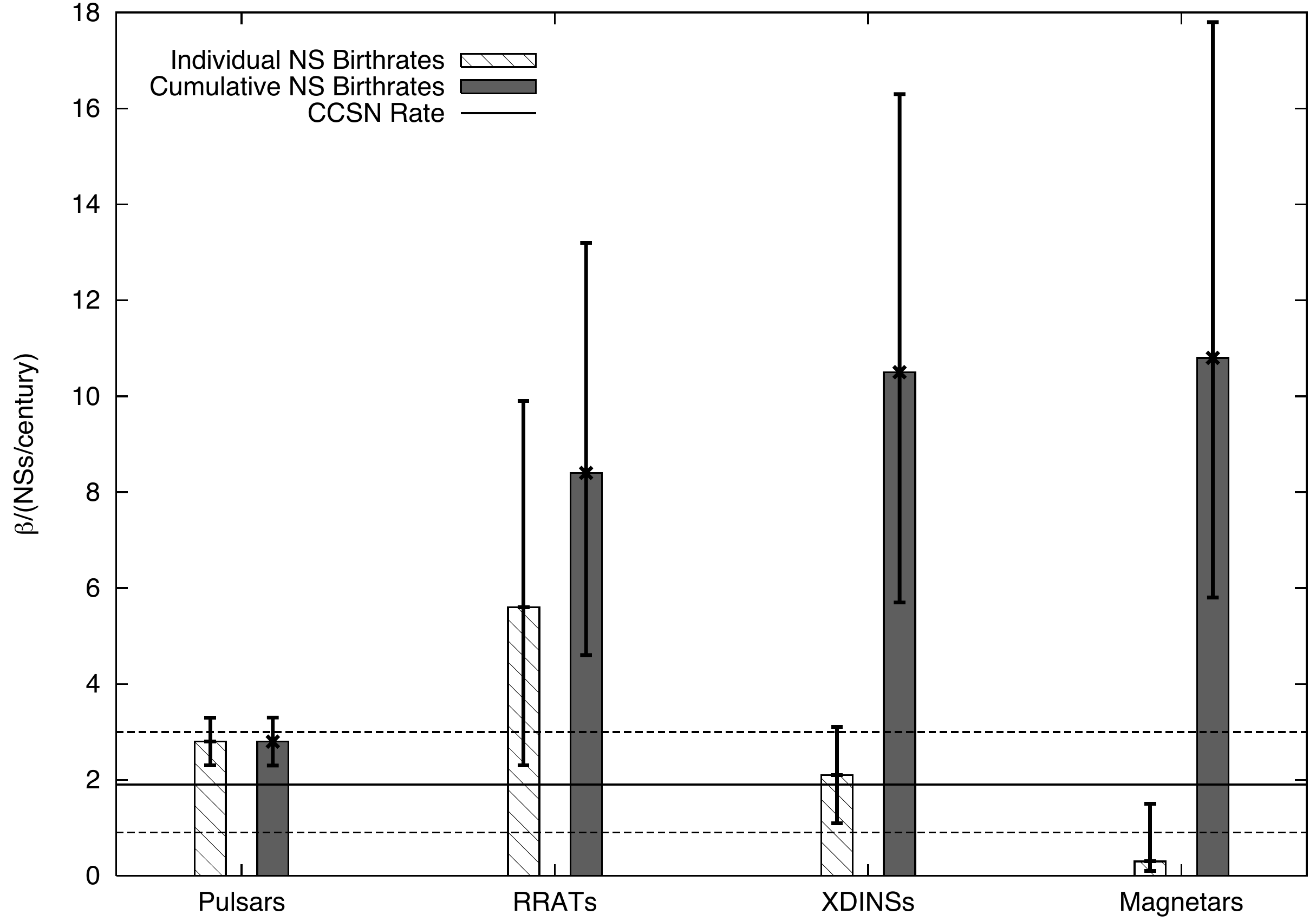}
	\end{center}
	\vspace{-10pt}
	\caption{Estimated birthrates for the individual NS populations and
          the cumulative number. The CCSN rate is shown as a horizontal line
          (solid), as are its error bars (dashed).}
	\label{fig:birthrates_plot1} 
	\vspace{-10pt}
\end{figure}

\section{Too Many Neutron Stars?}     % comparing the rates

The birthrates for each of the NS populations are summarised below in Table 1
and Figure 2. It appears that the CCSN rate cannot sustain all the separate NS
populations. In a previous consideration of this question by Popov et
al.~(2006)\nocite{ptp06}, XDINSs and RRATs were identified as a single NS
population, so that only one birthrate contribution was taken, i.e.~that of
the XDINSs. Moreover, the magnetar contribution was neglected and an XDINS
birthrate was assumed such that $\beta_{{\rm XDINS}}\approx\beta_{{\rm
    PSR}}$. In addition, Popov et al.~used the lower limit pulsar birthrate of
V+04 to be the pulsar birthrate, a result since superceded by the work of
FK06.  In this picture, where XDINSs are identified with nearby RRATs, the
total birthrate is $\beta_{{\rm total}}=2-4$/century which is barely
consistent with the CCSN rate.  However, including the magnetars, allowing for
separate RRAT and XDINS contributions and using the more accurate pulsar
birthrate of FK06, Eqn.~\eqref{balance_equation} cannot be satisfied with the
estimates from Table 1. This seems to be the case even if we assume the
highest CCSN rate allowable within the uncertainties
(i.e.~$\beta_{CCSN}=3$/century) while at the same time allowing for the lowest
required total required NS birthrate, $\beta_{total}=5.8$/century. It seems
that the number of NSs  produced via CCSNe is not sufficient.

We can just about reconcile the rates if we choose the highest allowable CCSN
rate and the lowest allowable total NS birthrate from the L+06 result using
the TC93 electron density model (see Table 1).  However, as we discussed
earlier, the pulsar current results are lower limits and the NE2001 model is
often considered to be a more accurate model than TC93 \citep{cl02,cl03},

From looking at Figure 2 we are left to conclude that either the individual 
NS birthrates are over-estimated (or the uncertainties in these values are 
under-estimated). To reconcile the values within the errors would require the 
RRAT and XDINSs errors (which recall are the most uncertain) to be 
under-estimated by a factor of 2. If this is not the case then it would seem that 
Eqn.~\eqref{balance_equation} is not satisfied. Taking this at face value implies 
that there are too many NSs in the Galaxy. We will discuss the nature of 
this potential NS ``birthrate problem'' in the following.

\section{Discussion}

In trying to determine some possible solutions to the birthrate problem we
consider in the following the possibility that the various birthrates are
incorrect or that there is an evolutionary answer. Some possible conclusions
include:

{\em (1) The Pulsar Birthrate is wrong:} The pulsar birthrate is the most
crucial component of our discussion as pulsars are the most well studied
population and the RRAT birthrate depends on that of the pulsars. Thankfully,
the pulsar birthrate estimates are by far the most accurate. The pulsar
current analyses make no assumptions and are ``model free'' even though they
depend on the Galactic electron density distribution and the beaming
fraction. The lower limits obtained from them are thus quite secure. In order to
compensate for the flux limited nature of these studies, we would need to to 
choose a functional form for the luminosity (depending on $P$ and
$\dot{P}$) but the inclusion of such a correction can only increase the
determined birthrate.

The work of FK06 models this luminosity evolution across the $P-\dot{P}$
diagram as well as many other birth properties (modelled from the observed
pulsar population).  The analysis did assume magnetic dipole spin-down of
pulsars but allowed for magnetic field decay as well as drawing braking
indices from a uniform distribution in the range $n \in [1.4,3.0]$ (note that
the few measured braking indices are found to lie in the range $1.4-2.9$, see
Lyne \& Graham-Smith (2004), ~\citet{liv+07} and references therein).

Another uncertainty for pulsars is the beaming fraction. An indication of
this may be the recent discovery of a pulsar with an extremely small duty
cycle (Keith et al.~2008). This pulsar has a beaming fraction of just $0.04\%$
or only $0.14^o$ of longitude. Usually, we would expect the minimum pulse
width (for an orthogonal rotator) for this pulsar period of $P=91$ ms to be given by
$W_{{\rm min}}(h)\sim 8.2^\circ(h/10{\rm km})^{1/2}$ where $h$ is the emission
height and we have assumed $\beta$, the impact parameter, to be small. 
What is observed is a pulsar that is narrower by a factor of $\sim10$. It is possible that the pulse
represents a cut at the very edge of the conical beam but this seems to be at
odds with the two observed distinct components in the pulse profile (Keith et
al. 2008)\nocite{keith+08}. Pulsars with pulse widths this narrow therefore
raise the question whether or not our beaming fraction estimates are
accurate. If they are in fact over-estimates then there may be many more 
pulses which we do not see.

In summary, taking the pulsar current analysis to provide a reliable lower
limit it seems indeed reasonable to take a pulsar birthrate of $\beta_{{\rm
    PSR}} = 2$/century as being quite conservative when the many low
luminosity pulsars are included.

{\em (2) The RRAT birthrate is wrong and hugely over-estimated:} The RRAT
birthrate depends on the RRAT population estimate being correct. This is based
on assumptions that the Galactic distribution of RRATs follows that of
pulsars, on assumptions about the impact of man-made Radio Frequency
Interference (RFI) during the analysis of the PMPS data as well as on beaming
and Galactic electron distribution models used and on RRAT burst rate
estimates. The full expression for the number of RRATs includes a factor for
each of these input assumptions and is given by (McLaughlin et al. 2006) 
\begin{equation}
	N_{\rm RRAT}\approx2\times10^5 \left( \frac{100\; \mbox{\rm mJy kpc}^2} {L_{\rm min}} \right) \;
	\left( \frac{0.5}{f_{\rm on}} \right) \left( \frac{0.5}{f_{\rm RFI}} \right) \left(\frac{0.1}{f_{\rm beam}} \right),
\end{equation}
where $f_{{\rm on}}$ is the fraction of RRATs which were ``on'' during a
35-min PMPS observation, $f_{{\rm RFI}}$ is the fraction of RRATs not missed
due to RFI. $f_{{\rm beam}}$ is the beaming fraction 
where, due to recent studies, a modification to the equation has been made 
(Lorimer, private communication). All of these effects are treated conservatively but to really
improve the accuracy of the estimate, we need to discover many more sources.
This will enable us to accurately determine the factors in this equation.  The
``on''-factor can be best constrained with accurate RRAT burst rates for a
larger population of RRATs.  The RFI factor is more difficult to quantify but
recently we have made progress in this regard with the development of a new
RFI removal scheme which is capable of removing the vast majority of terrestrial
RFI (Eatough et al. in prep). The RFI removal means that we will be able to
essentially ignore the $f_{{\rm RFI}}$-factor (i.e. $f_{\rm
  RFI}\rightarrow1$). A re-processing of the PMPS data applying this superior RFI
scheme is underway and will eventually enable us to detect all RRATs that were
originally overlooked due to RFI contamination. This will yield a more
accurate value of $N_{{\rm RRAT}}$ and the results of this will be reported
later.

An uncertainty of the beaming fraction of radio pulsars also affects
RRATs as we have, as the best available assumption, adopted a beaming
fraction as observed for long period pulsars. If the beams are
narrower then this increases the estimate for the number of RRATs. The
RRAT lifetime is another uncertain parameter. If one were to propose a
much larger active lifetime than that proposed by Popov et al. (2006)
this would reduce the implied RRAT birthrate. However even if this
were an order of magnitude larger (i.e. $\sim50$ Myr) the birthrate
problem would remain although less emphatic. Conversely we note that
the RRATs seem to have higher $\dot{P}$ values than most pulsars and
thus may evolve to higher periods (i.e. towards the death valley) more
quickly than pulsars which would then imply a shorter lifetime than
that assumed above.  Noting all of these caveats, it does not seem
unreasonable to take, as a conservative estimate, $\beta_{{\rm
RRAT}}=2-6$/century for $\gamma=1-3$ as before.

{\em (3) The XDINS birthrate is wrong and hugely over-estimated:} Like in the
case of RRATs this could be due to the small sample size of this
population. Furthermore some of the XDINS birthrate estimates assume spin-down
ages to be an accurate age estimate and there are only three XDINSs with well
known $\dot{P}$. Furthermore, the estimate of~\cite{gh07} also depends on the
used Galactic $N_{{\rm HI}}$ model, i.e.~the authors applied an exponential
model (ignoring any warp or spiral arm components). Long-term monitoring of
the known XDINSs can yield exact period derivatives which in turn will give us
accurate characteristic ages which, along with NS cooling ages, may enable us
to determine the true age of XDINSs. Certainly, however, as for RRATs, the
best way to improve the used estimates is to increase the known population of
XDINSs. Recent work to determine where best to search for these elusive
sources, has pointed towards the Cygnus-Cepheus region behind the Gould Belt
(Posselt et al. 2008).\nocite{pph+08}

{\em (4) A possible evolution from different types of NSs into others:} We
consider here the possibility that pulsars, RRATs and XDINSs might be
different evolutionary stages of a single class of object. If this is the
case, we need only take one birthrate for these populations into balancing the
CCSN rate, i.e.~the birthrate of the earliest stage in this cycle. To
determine the direction of this evolution requires a reliable age
estimator. As all of the considered NSs are isolated objects, we expect that
they simply slow down as they age, so that the longer periods of XDINSs
($3-11$ s), compared with those of the RRATs ($0.7-7$ s), imply RRATs to be
younger than XDINSs. Similarly, the even lower periods of isolated pulsars
($0.03-8.5$ s) might imply that the evolutionary track in question is
pulsar$\rightarrow$RRAT$\rightarrow$XDINS.  To investigate this possibility,
we performed Kolmogorov-Smirnov (K-S) tests comparing the cumulative period
probability distributions (see Figure 3). The probability that the pulsar and
RRAT distributions are drawn from the same parent distribution is found to be
$0.02\%$. However, we believe the comparison to be unfair due to the large
difference in distribution sizes (1500 and 11).  To test this we randomly
selected 20 pulsar periods from their period distribution and compared these
with the RRAT distribution over many iterations. The resulting probabilities
vary largely with $\sim17\%$ of iterations showing probabilities below $1\%$
but $\sim18\%$ of iterations showing probabilities above $20\%$.  Next we
compared the median periods of the randomly selected pulsar samples to that of
the RRATs.  This value is stable with the median pulsar period being $615$ ms
over 10,000 iterations. The RRAT median period is $\sim7\sigma$ above this
value considering 16 known RRATs (the 11 original sources plus 5 newly
detected sources, M. McLaughlin, private communication) and $\sim20\sigma$
considering only the published sources. From this we conclude that RRAT
periods are intrinsically longer than those of the pulsars.

Comparing the RRAT and XDINS distributions using the K-S test gives a
probability that these two distributions are drawn from the same parent
distribution of only $2\%$ but not low enough to reject this possibility given
the small numbers in each category ($N_{{\rm RRAT}}=11$, $N_{{\rm
      XDINS}}=7$). If we nevertheless assume these distributions to be
  different, we can estimate the time needed for the RRAT period distribution
  to evolve to the XDINS period distribution by comparing the average period
  and the period derivatives.  The average periods are, respectively, $\langle
  P_{{\rm RRAT}}\rangle=3.6$ s and $\langle P_{{\rm XDINS}}\rangle=8.1$
  s. Using the average RRAT period derivative of
  $\langle\dot{P}\rangle=2\times10^{-13}$ and assuming that it is constant
  with time, we estimate an evolutionary time of
\begin{equation}
	t=\frac{\langle P_{{\rm XDINS}}\rangle-\langle P_{{\rm RRAT}}\rangle}{\langle\dot{P}_{{\rm RRAT}}\rangle}\sim0.7\; \mbox{\rm Myr.} \;
\end{equation}

\begin{figure}
	\vspace{-10pt}
	\begin{center}
		\includegraphics[scale=0.3]{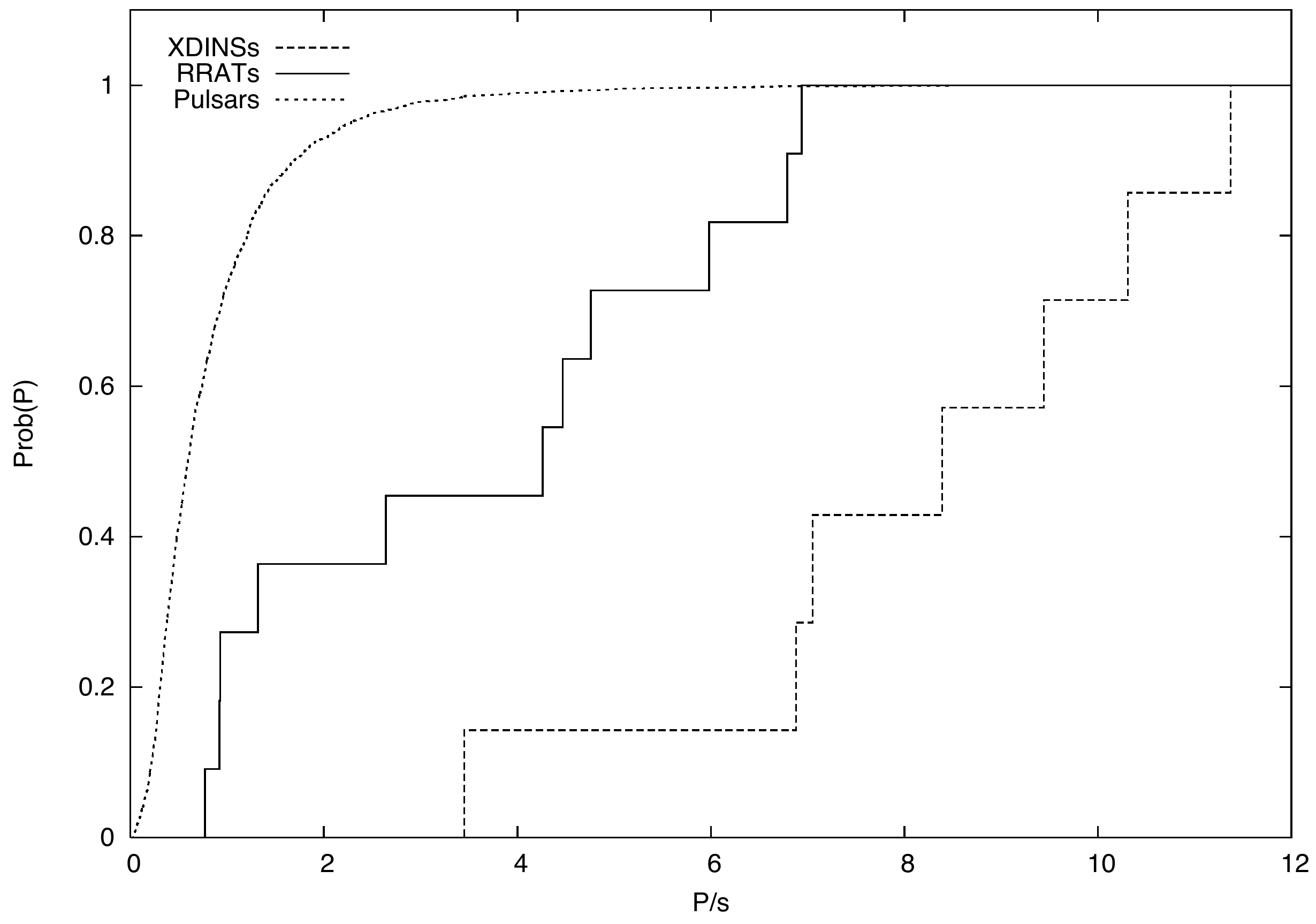}
	\end{center}
	\vspace{-15pt}
	\caption{\small{Cumulative probability distributions for pulsars, RRAT and XDINS periods.}}
	\vspace{-15pt}
\end{figure}

Another indicator of age is temperature, assuming that no significant heating
occurs during the life of any of these sources. Assuming NSs are born with the
same temperature and then cool along a NS cooling curve, we can determine the
age if the surface temperature can be measured\footnote{We note the extra
  downward emission above the surfaces of pulsars will act as a heating
  mechanism of the polar cap. This scenario does not apply for RRATs and
  XDINSs which are ``off'' most and all of the time}.  However, NS cooling
curves are not very well constrained (as they often depend on the unknown NS
equation-of-state, e.g.~Yakovlev \& Pethick 2004) so that the exact
age-temperature relationship is not known. Still, we can safely expect an
older star to be cooler than a young NS, so that we might suppose XDINSs to be
cooler than RRATs as XDINSs have ${\rm T\approx0.7\times10^{6}}$
K~\citep{yp04} and RRAT J1819$-$1458 has ${\rm T \approx1.6\times10^{6}}$ K,
lending tentative support to our conclusion based upon periods.

If some/all RRATs do evolve into XDINSs it means the XDINS birthrate can be
removed from consideration in balancing Eqn.~\eqref{balance_equation}. We note that
a NS evolution from RRAT to XDINS suggests a path on the $P-\dot{P}$ diagram
(along, say, an $n=3$ line) which starts in the region of the high-B radio
pulsars.  Such an evolution where RRATs and XDINSs are evolutionary states
reached by some of the normal radio pulsars means both of these birthrate
contributions can be removed from Eqn.~\eqref{balance_equation} and the birthrate
problem is much less severe and is solved within the errors of $\beta_{{\rm
    PSR}}$ and $\beta_{{\rm CCSN}}$.

This of course does not include the magnetar contribution. If the magnetar
birthrate is in fact small, it is not necessary to include magnetars in an
evolutionary scenario to solve the problem but nonetheless we consider
it. Comparing the magnetar period distribution to that of RRATs and XDINSs
suggests that the XDINSs and magnetar distributions are quite similar. A K-S
test finds their distributions to be the same with a $92\%$ probability. This
might suggest that the XDINS and magnetar evolutionary end states come from
the same initial population. Such a scenario with some high-B radio pulsars
evolving towards the magnetar region of the $P-\dot{P}$ diagram involves
increasing $B_{{\rm S}}$. While at first this may seem counter-intuitive we
note that Eqn.~(2) describes the surface magnetic field which could indeed
increase with time. In fact, this is suggested by long term observations of the Crab pulsar
and PSR B1737$-$30~\citep{ls04,lyn04}, and by the measured braking indices from young
pulsars as $n<3$.

{\em (5) An unknown NS formation process:} Another (even more drastic)
solution might be that there is some other unknown mechanism of forming NSs
besides CCSNe. This might well be required even if there is an evolutionary
answer to the birthrate problem because, as we have noted, the number of
pulsars alone is pushing the allowed limit. The only other known mechanisms
for forming NSs are electron-capture SNe~\citep{nomoto84,nomoto87}  and an
accretion induced collapse (AIC)~\citep{gri87}.
 
While stellar modelling results~\citep{et04,pnp+04} had seen electron-capture SNe 
only possible in close binary systems\footnote{It is thought that the lower mass pulsar in the 
double pulsar system, J0737$-$3039B, was formed in this way.}, some 
very recent work has examined this process in isolated stars (Poelarends et al. 2008).\nocite{phlp08} 
The results suggest that electron capture SNe likely account for just $\sim4 \%$ of all 
SNe (i.e. not just CC SNe) which increases the number of Galactic NSs but only by a small amount.
  
AIC is a process naturally only occurring in binary systems but here we consider 
only isolated NSs so this process cannot resolve the problem. It would seem that another
as yet unknown NS formation mechanism would be required.

\section{Conclusions}

In this work we have presented a detailed and critical review of the current
state of birthrate calculations for a number of NS manifestions. Based on this
review, we suggest that the current CCSN rate cannot explain the birthrates of
the various NS populations. Unless new birthrate estimates emerge which differ 
vastly from those discussed here we have a birthrate problem. If this is the case 
we favour an evolutionary interpretation where radio pulsars, RRATs and XDINSs 
(and possibly also magnetars) are different evolutionary stages of the same object. 
Another possible, more exotic solution would be the existence of some, as yet unknown, 
mechanism for forming NSs. While the determined birthrates are uncertain for XDINSs, RRATs
and magnetars we consider the currently claimed uncertainties, if in fact they are correct, 
not to be large enough to convincingly remove the described problem. 

To truly advance in answering these questions,
it is essential to find a significantly larger number of sources to increase
the known populations of RRATs, XDINSs and magnetars. Searches to this end are
underway and many more are planned. There are prospects for 
discovering XDINSs and magnetars from future high energy observatories. These include the 
International X-Ray Observatory (which has now superceded the planned 
XEUS and Con-X missions) as well as the recently launched Fermi 
gamma-ray space observatory. Radio surveys include the ongoing P-ALFA
survey at Arecibo (Cordes et al. 2006)\nocite{cordes+06}, the planned transients and pulsar observations with LOFAR, 
and the new Parkes and Effelsberg multi-beam all-sky surveys (beginning Autumn 2008).
Ultimately, a Galactic census of pulsars with the Square Kilometre Array (SKA)
should improve our knowledge of Galactic neutron stars phenomenally with the
expected detection of $\sim20000$ new pulsars \citep{ckl+04}. Together, the SKA 
and LOFAR will monitor the Northern and Southern hemisphere essentially continuously. 
Even though the spectrum of RRATs and hence their discovery potential at low frequencies 
still has to be assessed, both telescopes should allow us to potentially observe most RRATs 
in the Galaxy that are beaming towards Earth. In other words, SKA observations will, with 
certainty, establish the relative population numbers for RRATs and pulsars, confirming or 
rejecting the results of this study. Extremely valuable information will, however, be already 
available much sooner with the help of LOFAR (van Leeuwen \& Stappers 2008)~\nocite{ls08}.

\section*{Acknowledgements}

We would like to thank Sir Francis Graham-Smith, Duncan Lorimer, Maura McLaughlin 
and the anonymous referee for comments which have improved the quality of this work. 
EK acknowledges the support of a Marie-Curie EST Fellowship with the FP6 
Network ``ESTRELA" under contract number MEST-CT-2005-19669.
% THE END

%%% comment out these lines and include .bbl file in .tex file in FINAL version %%%
%\bibliographystyle{mn2e}
%\bibliography{psrrefs,modrefs}

\begin{thebibliography}{}

\bibitem[\protect\citeauthoryear{Alpar, Cheng, Ruderman \& Shaham}{Alpar
  et~al.}{1982}]{acrs82}
Alpar M.~A.,  Cheng A.~F.,  Ruderman M.~A.,    Shaham J.,  1982, {Nature}, 300,
  728

\bibitem[\protect\citeauthoryear{{Arras}, {Cumming} \& {Thompson}}{{Arras}
  et~al.}{2004}]{act04}
{Arras} P.,  {Cumming} A.,    {Thompson} C.,  2004, {ApJL}, 608, L49

\bibitem[\protect\citeauthoryear{{Camilo}, {Ransom}, {Halpern}, {Reynolds},
  {Helfand}, {Zimmerman} \& {Sarkissian}}{{Camilo} et~al.}{2006}]{cmh+06}
{Camilo} F.,  {Ransom} S.~M.,  {Halpern} J.~P.,  {Reynolds} J.,  {Helfand}
  D.~J.,  {Zimmerman} N.,    {Sarkissian} J.,  2006, {Nature}, 442, 892

\bibitem[\protect\citeauthoryear{{Camilo}, {Reynolds}, {Johnston}, {Halpern} \&
  {Ransom}}{{Camilo} et~al.}{2008}]{crj+08}
{Camilo} F.,  {Reynolds} J.,  {Johnston} S.,  {Halpern} J.~P.,    {Ransom}
  S.~M.,  2008, {ApJ}, pp 681--686

\bibitem[\protect\citeauthoryear{Chen \& Ruderman}{Chen \&
  Ruderman}{1993}]{cr93a}
Chen K.,  Ruderman M.,  1993, {ApJ}, 402, 264

\bibitem[\protect\citeauthoryear{Colpi, Geppert \& Page}{Colpi
  et~al.}{2000}]{cgp00}
Colpi M.,  Geppert U.,    Page D.,  2000, {ApJL}, 529, L29

\bibitem[\protect\citeauthoryear{Cordes, Kramer, Lazio, Stappers, Backer \& Johnston}{Cordes
  et~al.}{2004}]{ckl+04} Cordes J. M., Kramer, M., Lazio, T. J. W., Stappers, B. W. \& Johnston, S., 
  2004, {New Astronomy Reviews}, 48, 1413 - 1438

\bibitem[\protect\citeauthoryear{{Cordes} \& {Lazio}}{{Cordes} \&
  {Lazio}}{2002}]{cl02}
{Cordes} J.~M.,  {Lazio} T.~J.~W.,  2002, {ArXiv e-prints (astro-ph/0207156)}

\bibitem[\protect\citeauthoryear{{Cordes} \& Lazio}{{Cordes} \&
  Lazio}{2003}]{cl03}
{Cordes} J.~M.,  Lazio T.~J.~W.,  2003, {ArXiv e-prints (astro-ph/0301598)}

\bibitem[\protect\citeauthoryear{{{Cordes}, J. M. et al.}}{{{Cordes}, J. M. et
  al.}}{2006}]{cordes+06}
{{Cordes}, J. M. et al.} 2006, {ApJ}, 637, 446

\bibitem[\protect\citeauthoryear{{{Diehl}, D. et al.}}{{{Diehl}, D. et
  al.}}{2006}]{diehl+06}
{{Diehl}, D. et al.} 2006, Nature, 439, 45

\bibitem[\protect\citeauthoryear{{Eldridge} \& {Tout}}{{Eldridge} \&
  {Tout}}{2004}]{et04}
{Eldridge} J.~J.,  {Tout} C.~A.,  2004, {MNRAS}, 353, 87

\bibitem[\protect\citeauthoryear{Faucher-Giguere \& Kaspi}{Faucher-Giguere \&
  Kaspi}{2006}]{fk06}
Faucher-Giguere C.-A.,  Kaspi V.~M.,  2006, ApJ, 643, 332

\bibitem[\protect\citeauthoryear{{Gill} \& {Heyl}}{{Gill} \&
  {Heyl}}{2007}]{gh07}
{Gill} R.,  {Heyl} J.,  2007, MNRAS, 382, 52

\bibitem[\protect\citeauthoryear{Grindlay}{1987}]{gri87}
  Grindlay J.~E., 1987, in The Origin of Neutron Stars in Globular Clusters, 
  {IAU} Symposium {N}o. 113.
  eds.~D. J.\ Helfand \& J. H..\ Huang, Reidel, 
  Dordrecht, p.\ 173 - 185

\bibitem[\protect\citeauthoryear{Haberl}{Haberl}{2004}]{hab04}
Haberl F.,  2004, Adv. Space Res., 33, 638

\bibitem[\protect\citeauthoryear{{Haberl}}{{Haberl}}{2007}]{hab07}
{Haberl} F.,  2007, Astrophysics and Space Science, 308, 171

\bibitem[\protect\citeauthoryear{{Hankins}, {Kern}, {Weatherall} \&
  {Eilek}}{{Hankins} et~al.}{2003}]{hkwe03}
{Hankins} T.~H.,  {Kern} J.~S.,  {Weatherall} J.~C.,    {Eilek} J.~A.,  2003,
  Nature, 422, 141

\bibitem[\protect\citeauthoryear{{Hankins} \& {Rickett}}{1975}]{hr75}
  Hankins T.~H. \& Rickett B.~J., 1975, in Methods in Computational Physics 14, 
  eds.~B.\ Alder, S.\ Fernbach \& W.\ Rotenberg, Academic Press, 
  New York, p.\ 55 - 129

\bibitem[\protect\citeauthoryear{{Heger}, {Fryer}, {Woosley}, {Langer} \&
  {Hartmann}}{{Heger} et~al.}{2003}]{hfw03}
{Heger} A.,  {Fryer} C.~L.,  {Woosley} S.~E.,  {Langer} N.,    {Hartmann}
  D.~H.,  2003, {ApJ}, 591, 288

\bibitem[\protect\citeauthoryear{{Keith}, {Johnston}, {Kramer}, {Weltevrede},
  {Watters} \& {Stappers}}{{Keith} et~al.}{2008}]{keith+08}
{Keith} M.~J.,  {Johnston} S.,  {Kramer} M.,  {Weltevrede} P.,  {Watters} K.,
   {Stappers} B.,  2008, {ArXiv e-prints (astro-ph/0807.2088)}

\bibitem[\protect\citeauthoryear{{van Kerkwijk} \& {Kaplan}}{{van Kerkwijk} \&
  {Kaplan}}{2008}]{kk08}
{van Kerkwijk} M.~H.,  {Kaplan} D.~L.,  2008, ApJ, 673, L163

\bibitem[\protect\citeauthoryear{{Kondratiev}, {Burgay}, {Possenti},
  {McLaughlin}, {Lorimer}, {Turolla}, {Popov} \& {Zane}}{{Kondratiev}
  et~al.}{2008}]{kba+08}
{Kondratiev} V.~I.,  {Burgay} M.,  {Possenti} A.,  {McLaughlin} M.~A.,
  {Lorimer} D.~R.,  {Turolla} R.,  {Popov} S.,    {Zane} S.,  2008, ApJ,
  submitted

\bibitem[\protect\citeauthoryear{{Kouveliotou, C. et al.}}{{Kouveliotou, C. et
  al.}}{1998}]{kds+98}
{Kouveliotou, C. et al.} 1998, {Nature}, 393, 235

\bibitem[\protect\citeauthoryear{Kroupa}{Kroupa}{2001}]{kroupa01}
Kroupa P.,  2001, MNRAS, 322, 231

\bibitem[\protect\citeauthoryear{Kroupa}{Kroupa}{2002}]{kro02b}
Kroupa P.,  2002, Science, 295, 82

\bibitem[\protect\citeauthoryear{{van Leeuwen and Kuiper}}
{{van Leeuwen and Stappers}}{2008}]{ls08}
{van Leeuwen} J., {Stappers} B. W.,  2008, AIPC Proceedings, 983, 598

\bibitem[\protect\citeauthoryear{{Livingstone}, {Kaspi}, {Gavriil},
  {Manchester}, {Gotthelf} \& {Kuiper}}{{Livingstone} et~al.}{2007}]{liv+07}
{Livingstone} M.~A.,  {Kaspi} V.~M.,  {Gavriil} F.~P.,  {Manchester} R.~N.,
  {Gotthelf} E. V.~G.,    {Kuiper} L.,  2007, Ap\&SS, 308, 317L

\bibitem[\protect\citeauthoryear{{Lorimer, D. R. and Kramer, M.}}{{Lorimer, D.
  R. and Kramer, M.}}{2005}]{lk05}
{Lorimer, D. R. and Kramer, M.} 2005, {Handbook of Pulsar Astronomy}.
Cambridge University Press

\bibitem[\protect\citeauthoryear{{Lorimer, D. R. et al.}}{{Lorimer, D. R. et
  al.}}{2006}]{lfl+06}
{Lorimer, D. R. et al.} 2006, {MNRAS}, 372, 777

\bibitem[\protect\citeauthoryear{Lyne}{2004}]{lyn04}
  Lyne A.~G., 2004, in Young Neutron Stars and Their Environments, 
  {IAU} Symposium {N}o. 218.
  eds.~F.\ Camilo \& B. M.\ Gaensler, San Franciso, CA, p.\ 257

\bibitem[\protect\citeauthoryear{Lyne \& Smith}{Lyne \& Smith}{2004}]{ls04}
Lyne A.~G.,  Smith F.~G.,  2004, Pulsar Astronomy, 3rd ed.
Cambridge University Press, Cambridge

\bibitem[\protect\citeauthoryear{Malofeev, Gil, Jessner, Malov, Seiradakis,
  Sieber \& Wielebinski}{Malofeev et~al.}{1994}]{mgj+94}
Malofeev V.~M.,  Gil J.~A.,  Jessner A.,  Malov I.~F.,  Seiradakis J.~H.,
  Sieber W.,    Wielebinski R.,  1994, {A\&A}, 285, 201

\bibitem[\protect\citeauthoryear{{Manchester, R. N. et al.}}{{Manchester, R. N.
  et al.}}{2001}]{mlc+01}
{Manchester, R. N. et al.} 2001, MNRAS, 328, 17

\bibitem[\protect\citeauthoryear{{Maron}, {Kijak}, {Kramer} \&
  {Wielebinski}}{{Maron} et~al.}{2000}]{mkk+00}
{Maron} O.,  {Kijak} J.,  {Kramer} M.,    {Wielebinski} R.,  2000, {AASS}, 147,
  195

\bibitem[\protect\citeauthoryear{{McKee} \& {Ostriker}}{{McKee} \&
  {Ostriker}}{2007}]{mo07}
{McKee} C.~F.,  {Ostriker} E.~C.,  2007, {ArXiv e-prints (astro-ph/0707.3514)}

\bibitem[\protect\citeauthoryear{{{McLaughlin}, M.~A. et al.}}{{{McLaughlin},
  M.~A. et al.}}{2006}]{mll+06}
{{McLaughlin}, M.~A. et al.} 2006, Nature, 439, 817

\bibitem[\protect\citeauthoryear{{{McLaughlin}, M. A. et al.}}{{{McLaughlin},
  M. A. et al.}}{2007}]{mrg+07}
{{McLaughlin}, M. A. et al.} 2007, {ApJ}, 670, 1307

\bibitem[\protect\citeauthoryear{{Muno}, {Gaensler}, {Nechita}, {Miller} \&
  {Slane}}{{Muno} et~al.}{2008}]{muno+08}
{Muno} M.~P.,  {Gaensler} B.~M.,  {Nechita} A.,  {Miller} J.~M.,    {Slane}
  P.~O.,  2008, ApJ, 680, 639

\bibitem[\protect\citeauthoryear{{Nomoto}}{{Nomoto}}{1984}]{nomoto84}
{Nomoto} K.,  1984, {ApJ}, 277, 791

\bibitem[\protect\citeauthoryear{{Nomoto}}{{Nomoto}}{1987}]{nomoto87}
{Nomoto} K.,  1987, {ApJ}, 322, 206

\bibitem[\protect\citeauthoryear{{van Paradijs}, Taam \& {van den Heuvel}}{{van
  Paradijs} et~al.}{1995}]{vtv95}
{van Paradijs} J.,  Taam R.~E.,    {van den Heuvel} E. P.~J.,  1995, {A\&A},
  299, L41

\bibitem[\protect\citeauthoryear{Phinney \& Blandford}{Phinney \&
  Blandford}{1981}]{pb81}
Phinney E.~S.,  Blandford R.~D.,  1981, {MNRAS}, 194, 137

\bibitem[\protect\citeauthoryear{Podsiadlowski, Langer, Poelarends, Rappaport,
  Heger \& Pfahl}{Podsiadlowski et~al.}{2004}]{pnp+04}
Podsiadlowski P.,  Langer N.,  Poelarends A. J.~T.,  Rappaport S.,  Heger A.,
   Pfahl E.~D.,  2004, {ApJ}, 612, 1044
   
\bibitem[\protect\citeauthoryear{Poelarends, Herwig, Langer \& Pfahl}
{Podsiadlowski et~al.}{2004}]{phlp+04}
Poelarends A.J.~T., Herwig, F., Langer N., Heger A.,  
2008, {ApJ}, 675, 614
      
\bibitem[\protect\citeauthoryear{Popov, Turolla \& Possenti}{Popov
  et~al.}{2006}]{ptp06}
Popov S.~B.,  Turolla R.,    Possenti A.,  2006, MNRAS, 369, L23

\bibitem[\protect\citeauthoryear{{Poppel}}{{Poppel}}{1997}]{poppel97}
{Poppel} W.,  1997, Fund. Cos. Phys., 18, 1

\bibitem[\protect\citeauthoryear{{Posselt}, {Popov}, {Haberl}, {Truemper},
  {Turolla} \& {Neuhaeuser}}{{Posselt} et~al.}{2008}]{pph+08}
{Posselt} B.,  {Popov} S.~B.,  {Haberl} F.,  {Truemper} J.,  {Turolla} R.,
  {Neuhaeuser} R.,  2008, {A\&A}, 482, 617

\bibitem[\protect\citeauthoryear{{{Reynolds}, S. M. et al.}}{{{Reynolds}, S. M.
  et al.}}{2006}]{reyn+06}
{{Reynolds}, S. M. et al.} 2006, {ApJ}, 639, L71

\bibitem[\protect\citeauthoryear{Stahler \& Palla}{Stahler \&
  Palla}{2004}]{sp04}
Stahler W.~W.,  Palla F.,  2004, {The Formation of Stars}.
{Wiley-VCH}, {Weinheim, Germany}

\bibitem[\protect\citeauthoryear{Tauris \& Manchester}{Tauris \&
  Manchester}{1998}]{tm98}
Tauris T.~M.,  Manchester R.~N.,  1998, MNRAS, 298, 625

\bibitem[\protect\citeauthoryear{Taylor \& Cordes}{Taylor \&
  Cordes}{1993}]{tc93}
Taylor J.~H.,  Cordes J.~M.,  1993, {ApJ}, 411, 674

\bibitem[\protect\citeauthoryear{{Tiengo} \& {Mereghetti}}{{Tiengo} \&
  {Mereghetti}}{2007}]{tm07}
{Tiengo} A.,  {Mereghetti} S.,  2007, {ApJ}, 657, L101

\bibitem[\protect\citeauthoryear{Vivekanand \& Narayan}{Vivekanand \&
  Narayan}{1981}]{vn81}
Vivekanand M.,  Narayan R.,  1981, {JApA}, 2, 315

\bibitem[\protect\citeauthoryear{{{Voges}, W. et al.}}{{{Voges}, W. et
  al.}}{1999}]{voges+99}
{{Voges}, W. et al.} 1999, {A\&A}, 349, 389

\bibitem[\protect\citeauthoryear{{{Vranesevic}, N. et al.}}{{{Vranesevic}, N.
  et al.}}{2004}]{vml+04}
{{Vranesevic}, N. et al.} 2004, ApJL, 617, L139

\bibitem[\protect\citeauthoryear{{Walter}, {Wolk} \& {Neuhauser}}{{Walter}
  et~al.}{1996}]{wwn96}
{Walter} F.~M.,  {Wolk} S.~J.,    {Neuhauser} R.,  1996, {Nature}, 379, 233

\bibitem[\protect\citeauthoryear{Woods \& Thompson}{2004}]{wt04}
Woods P.~M., \& Thompson C., 2004, in Compact Stellar X-ray Sources, eds.
  W.~H.~G. Lewin \& M.~van~der Klis (United Kingdom: Cambridge University
  Press), in press (astro-ph/0406133)
  
\bibitem[\protect\citeauthoryear{{Yakovlev} \& {Pethick}}{{Yakovlev} \&
  {Pethick}}{2004}]{yp04}
{Yakovlev} D.~G.,  {Pethick} C.~J.,  2004, {ARA\&A}, 42, 169

\end{thebibliography}

\end{document}